\documentclass[preprint]{revtex4}
\usepackage{graphicx,epsfig,rotate} 
\def\iso#1#2{\mbox{${}^{#2}{\rm #1}$}}
\def\he#1{\iso{He}{#1}}
\def\li#1{\iso{Li}{#1}}
\def\be#1{\iso{Be}{#1}}

\def\ion#1#2{\mbox{#1$\;${\small\rmfamily #2}\relax}}

\def\ga{\mathrel{\raise.3ex\hbox{$>$\kern-.75em\lower1ex\hbox{$\sim$}}}}
\def\la{\mathrel{\raise.3ex\hbox{$<$\kern-.75em\lower1ex\hbox{$\sim$}}}}

\def\beq{\begin{equation}}
\def\eeq{\end{equation}}

\begin{document}

\rightline{UMN--TH--2714/08}
\rightline{FTPI--MINN--08/32}
\rightline{August 2008}

\title{A Bitter Pill:  The Primordial Lithium Problem Worsens}

\author{Richard H. Cyburt}
\affiliation{Joint Institute for Nuclear Astrophysics (JINA), National Superconducting Cyclotron Laboratory (NSCL), Michigan State University, East Lansing, MI 48824}
\author{Brian D. Fields}
\affiliation{Departments of Astronomy and of Physics, University of Illinois,
Urbana, IL 61801}
\author{Keith A. Olive}
\affiliation{William I. Fine Theoretical Physics Institute, 
University of Minnesota, Minneapolis, MN 55455, USA}

\begin{abstract}
The lithium problem arises from the significant discrepancy between
the primordial \li7 abundance as predicted by BBN theory and the WMAP
baryon density, and the pre-Galactic lithium abundance inferred from
observations of metal-poor (Population II) stars.  This problem has
loomed for the past decade, with a persistent
discrepancy of a factor of $2-3$ in
\li7/H.  Recent developments have sharpened all aspects of the Li
problem.  Namely: (1) BBN theory predictions have sharpened due to new
nuclear data, particularly the uncertainty on
$\he3(\alpha,\gamma)\be7$, has reduced to 7.4\%, and with a central
value shift of $\sim +$0.04 keV barn.  (2) The WMAP 5-year data now
yields a cosmic baryon density with an uncertainty reduced to 2.7\%.
(3) Observations of metal-poor stars have tested for systematic
effects, and have reaped new lithium isotopic data.  With these, we
now find that the BBN+WMAP predicts $\li7/{\rm H} = (5.24^{+0.71}_{-0.67})
\times 10^{-10}$.  The Li problem remains and indeed is exacerbated;
the discrepancy is now a factor 2.4 -- 4.3 or $4.2 \sigma$ (from
globular cluster stars) to $5.3\sigma$ (from halo field
stars). Possible resolutions to the lithium problem are briefly
reviewed, and key nuclear, particle, and astronomical measurements
highlighted.
\end{abstract}

\pacs{}
\keywords{}

\maketitle

\section{Introduction}

Big bang nucleosynthesis (BBN) remains one of few probes of the early
Universe with direct experimental or observational consequences
\cite{bbn}.  Historically, BBN was generally taken to be a
three-parameter theory with results depending on the baryon density of
the Universe, the neutron mean-life, and number of neutrino flavors.
Indeed, concordance between theory and observation for the abundances
of the light elements, D, \he3, \he4, and \li7 was a powerful tool for
obtaining the baryon density of the Universe. Over the last twenty or
so years, the number of light neutrino flavors has been fixed (in the
Standard Model of particle physics) \cite{pdg08}, and the neutron mean-life is measured
accurately enough that it is no longer a parameter, but rather its
residual uncertainty is simply carried into primarily an uncertainty
in the predicted \he4 abundance (and is small) \cite{pdg08}.

More recently, the baryon density has been determined to unprecedented
accuracy by WMAP \cite{wmap} to be $\Omega_B h^2 = 0.02273 \pm
0.00062$ where $\Omega_B = \rho_B/\rho_c$ is the fraction of critical
density in baryons, $\rho_c = 1.88 \times 10^{-29} h^2$ g cm$^{-3}$,
and $h$ is the Hubble parameter scaled to 100 km/Mpc/s. This
corresponds to a baryon-photon ratio of $\eta_{10} = 6.23 \pm 0.17$
where $\eta = n_B/n_\gamma = 10^{-10} \eta_{10}$.  This is
significantly more accurate than any determination of $\eta$ from
observational determinations of light element abundances.

Thus the paradigm for BBN has shifted in the post-WMAP era
\cite{cfo2,cfo3}.  Nucleosynthesis is now a parameter-free theory. In
the Standard Model, and working in the framework of a
Friedmann-Robertson-Walker cosmology, the only inputs are the nuclear
reaction rates (and their associated uncertainties). The last
significant update to many of the needed rates was compiled by the
NACRE group \cite{nacre} and recent BBN calculations by several groups
are in good agreement \cite{cfo1,cvcr01, coc,cyburt, coc2,cuoco}.

What has emerged is rather excellent agreement between the predicted
abundance of D/H as compared with the determined abundance from quasar
absorption systems \cite{pettini}.  Indeed what is often termed the
success in cosmology between BBN and the CMB is in reality only the
concordance between theory and observation for D/H at the WMAP value
of $\eta$.  Currently, there is no discrepancy between theory and
observation for \he4.  But this success is tempered by the fact that
\he4 is a poor baryometer, it varies only logarithmically with $\eta$,
and the observational uncertainty it rather large \cite{osk,fk}.

It has also become generally accepted that there is a problem
concerning the abundance of \li7.  At the WMAP value of $\eta$, the
predicted abundance of \li7 is approximately 3 times the
observationally determined value.  Several attempts at explaining this
discrepancy by adjusting some of the key nuclear rates proved
unsuccessful \cite{coc,coc3,cfo4}.  In fact, the key process for the
production of \li7 (in the form of \be7) is \he3($\alpha,\gamma$)\be7
and has been remeasured recently by several groups \cite{he3alpha} and one
of the purposes of this paper is to explore the consequences of these
new rates on the BBN production of \li7.  At this time, it is unclear
whether or not any known systematic effect may be responsible for the
discrepancy or whether there is an indication of new physics at play.

In what follows, we will review the status of the
\he3($\alpha,\gamma$)\be7 reaction rate and employ a new thermal fit
to this rate.  We will also briefly review the status of the \li7
observations as well as that of the other light elements.  In section
III, we will present the results of the BBN calculations using the new
\he3($\alpha,\gamma$)\be7 rates and quantify the discrepancy with observations.
In section IV, we discuss the remaining known alternative explanations
of the discrepancy and we give our conclusions in section V.

\section{New Light on the Lithium Problem}

\subsection{BBN Theory:  Updated Nuclear Data}

The possibility of systematic errors in the $\he3(\alpha,\gamma)\be7$
reaction, which is the most important \li7 production process in BBN,
was considered in detail in \cite{cfo4}.  The absolute value of the
cross section for this reaction is known relatively poorly both
experimentally and theoretically.  However, the agreement between the
standard solar model and solar neutrino data provides additional
constraints on possible systematic shifts in this cross section.
Using the standard solar model of Bahcall \cite{bah}, and recent solar
neutrino data \cite{sno}, one can exclude systematic variations of the
magnitude needed to resolve the BBN \li7 problem at the $\ga 95\%$
confidence level \cite{cfo4}.

In order to maintain up-to-date nuclear input for BBN, several new
efforts need to be addressed.  In particular, the recent experiments
measuring the neutron lifetime, evaluations of the $p(n,\gamma)d$ and
$\he3(\alpha,\gamma)\be7$ data, and the resulting revisions to
the $\he3(\alpha,\gamma)\be7$ cross
section and its error budget. 

The new neutron lifetime measurement by~\cite{serebrov} finds a
lifetime, $\tau_n=878.5\pm0.8$.  This is more than $5\sigma$ away from
the current world average
$\tau_n=885.7\pm0.8$  recommended by the Particle Data Group~\cite{pdg08}.  
Obviously, this points to unknown
systematics between the experiments.  Several possible explanations
exist~\cite{lamoreaux}, but remain to be explored experimentally.  We
thus, follow the PDG'08 recommendation, 
until some understanding of the discrepancy can be found.

The $p(n,\gamma)d$ determines at what point the deuterium bottleneck
ends.  Experimentally, this reaction is quite difficult to measure in
the relevant energy range for BBN.  We thus rely on theoretical-based
cross-sections which are normalized to the existing experimental data.
Several methods have been used to determine the shape of this cross
section; microscopic potential models~\cite{nakamura}, $R$-matrix
fits~\cite{hale}, N$^4$LO pion-less effective field
theory~\cite{rupak}, and NLO di-baryon effective field
theory~\cite{ando}.  All methods agree well, though the $R$-matrix fit
deviates by up to $\sim4$\% from the others.  Evaluating all the
relevant data, $np$-scattering, deuteron ground state properties as well
as the radiative capture and photo-dissociation data using a Markov
Chain Monte Carlo algorithm,  a best fit with propagated
uncertainties~\cite{ando}, finds $\la 1$\% errors in the capture rate.
We adopt this rate in our network.  The new rate is $\sim$1\% higher
than that adopted by~\cite{cyburt} at $E \sim 500$ keV.

At $\eta_{\rm WMAP}$, \li7 is produced as \be7, and the only important
reactions are production via $\he3(\alpha,\gamma)\be7$ and destruction
via $\be7(n,p)\li7$ and subsequent $\li7(p,\alpha)\he4$.  Of these,
new developments in $\he3(\alpha,\gamma)\be7$ experimental
efforts~\cite{he3alpha}, must be taken into account.  The apparent
discrepancy between prompt gamma-ray and \be7 activation measurements
discussed in~\cite{adelberger} has virtually disappeared, though
discrepancies between datasets still persist.  A new
evaluation~\cite{cybdav}, examines the ``modern'' data and finds a
best fit and 68.3\% CL uncertainties for a model independent fit, using a
Markov Chain Monte Carlo algorithm similar to that of~\cite{ando}.
Special care is taken to properly propagate the dominating
systematic normalization errors, as standard statistical treatments
underestimate the true uncertainties.  They find a zero-energy
astrophysical $S$ factor $S_{34}(0)=0.580\pm0.043$ keV b,
an uncertainty of 7.4\%.  In the
Gamow window ($E \sim 200$ keV), this new fit gives
a central value 17\% higher than
previous work~\cite{cyburt}.

Using the scalings provided in~\cite{cyburt} (their Eqn. 47), the two new
rates and baryon density, we expect an increase in the \li7 prediction to
$\sim5.3\times 10^{-10}$.  This increase combined with reduced errors will 
aggravate the Li problem as we show below.

\subsection{CMB Cosmological Parameters:  Updated Baryon Density}

As noted above, the major paradigm shift for BBN came with the first
WMAP determination of the baryon density.  The first year WMAP best
fit, assuming a variable spectral index was \cite{wmap1} $\Omega_B h^2
= 0.0224 \pm 0.0009$ which corresponds to a value of $\eta$:
$\eta_{10}(\mbox{WMAP2003}) = 6.14 \pm 0.25$.  This level of accuracy
was already able to pin the light element abundances down to a
narrow range.

The 5-year data WMAP data are consistent with their first year data,
and the errors have been significantly reduced.  The 5-year data give
$\Omega_B h^2 = 0.02273 \pm 0.00062$, or
\beq
\label{eq:eta_wmap}
\eta_{10}(\mbox{WMAP2008}) = 6.23 \pm 0.17
\eeq
We will adopt this value for the present study.

It is worth recalling key assumptions which enter into the inference of
$\eta$ from the CMB.  The baryon density arises from the pattern of
acoustic oscillations which modulate an underlying primordial spectrum
of temperature perturbations (e..g., \cite{hu}).  The baryon density
inferred from CMB data thus depends on the nature of the primordial
perturbation power spectrum, which is usually assumed (with motivation
from inflation) to be a simple or running power law, close to the
Harrison-Zel'dovich form.  If the primordial power spectrum had a
richer form, however, then the cosmological parameters inferred from
the temperature anisotropies can be changed substantially
\cite{blanchard}.  Fortunately, the inclusion of precision
polarization data, and their cross-correlation with the temperature
data, goes far to independently characterize the primordial spectrum
and to sharpen the precision of cosmic parameters.  The launch of {\em
Planck} in the next few months will offer just such precision
determination of CMB polarization, and thus should further reduce the
systematic and statistical uncertainties in the baryon density.

In using the WMAP value for $\eta$ at the period of BBN, we are
implicitly assuming that there has been no entropy change between BBN
and the decoupling of the CMB.  Note that entropy production between
BBN and decoupling would require a {\em larger} value for $\eta$ at
the time of BBN and make the Li problem even worse.

\subsection{Stellar Lithium Observations:  Updates, Systematics and Isotopic Ratios}
\label{sect:observations} 

There have been several recent observational determinations of the
\li7 abundance in metal-poor halo stars.  Most observations lead to a
\li7 abundance in the range $(1 - 2) \times 10^{-10}$, consistent with 
the original determination by Spite and Spite \cite{ss}. Precision
data have suggested a small but significant correlation between Li and
Fe \cite{rnb} which can be understood as the result of Li production
from Galactic cosmic rays \cite{fo,van}. Extrapolating to zero
metallicity one arrives at a primordial value \cite{rbofn} ${\rm
Li/H}|_p = (1.23 \pm 0.06) \times 10^{-10}$, though the systematic
uncertainties were recognized to be large and an abundance of
%%%$(1.23^{+0.68}_{-0.32}) \times 10^{-10}$ 
\beq
\label{eq:field}
({\rm Li/H})_{\rm p,field\star} = (1.23^{+0.68}_{-0.32}) \times 10^{-10}
\eeq
was derived at 95\% confidence.
 
There have been several measurements of \li7 abundances in globular
clusters.  
The abundance in NGC 6397 was measured by 
%%%three groups to be ${\rm Li/H}|_p = (2.19 \pm 0.28) \times 10^{-10}$ \cite{bon1},
\cite{bon1} to be 
\beq
\label{eq:gc}
({\rm Li/H})_{\rm p,GC} = (2.19 \pm 0.28) \times 10^{-10}
\eeq
Other groups found similar values of
$(1.91 \pm 0.44) \times 10^{-10}$~\cite{pm} and \li7/H = $(1.69 \pm
0.27) \times 10^{-10}$~\cite{thev}.  A related study (also of globular
cluster stars) gives \li7/H = $(2.29 \pm 0.94)
\times10^{-10}$~\cite{bon2}.

An important source for systematic error lies in the derived effective
temperature of the star.  [Li] $ = \log (\li7/{\rm H}) + 12$ is very
sensitive to the temperature, with $\partial {\rm [Li]}/\partial
T_{\rm eff} \simeq$ 0.065 -- 0.08.  Unfortunately there is no industry
standard for determining effective temperatures, and for a given star,
there is considerable range depending on the method used.  This spread
in temperatures was made manifest in the recent work of Melendez and
Ramirez \cite{mr} using the infra-red flux method (IRFM) which showed
differences for very low metallicities ([Fe/H] $<$ -3) by as much as
500 K, with typical differences of $\sim 200$ K with respect to that
of \cite{rnb}.  As a consequence the derived \li7 abundance was
significantly higher with ${\rm Li/H}|_p = (2.34 \pm 0.32) \times
10^{-10}$ \cite{mr,fov}.
 
Recent observations of metal-poor stars do not find support for the
large temperature scale in \cite{mr}.  Using H-$\alpha$ wings Asplund
et al. \cite{asp} found temperatures consistent with previous
determinations using the IRFM. This work also showed a non-negligible
correlation with metallicity leading to an estimate of the primordial
\li7 abundance between 1.1 and 1.5 $\times 10^{-10}$.  A larger slope
(with larger uncertainty) was also seen in \cite{bon3} (also using
H-$\alpha$ lines) in their study of extremely metal-poor stars.
Because of the large slope seen in this data, the extrapolation to
zero metallicity leads to a primordial \li7 abundance of 8.72 $\pm
1.71 \times 10^{-11}$, below that found in previous studies (though
their mean value is consistent with other determinations).
 
A dedicated set of observations were performed with the specific goal
of determining the effective temperature in metal-poor stars
\cite{hos}.  Using a large set of \ion{Fe}{I} excitation lines ($\sim 100$
lines per star), the Boltzmann equation was used with the excitation
energies, $\chi_i$ to determine the temperature through the
distribution of excited levels.  Again, there was no evidence for the
high temperatures reported in \cite{mr}, rather, temperatures were
found to be consistent with previous determinations.  The mean \li7
abundance found in \cite{hos} was ${\rm Li/H} = (1.3 - 1.4 \pm 0.2)
\times 10^{-10}$, consistent with the bulk of prior abundance
determinations.

There are of course other possible sources of systematic uncertainty
in the \li7 abundance.  It is possible that some of the surface \li7
has been depleted if the outer layers of the stars have been
transported deep enough into the interior, and/or mixed with material
from the hot interior; this may occur due to convection, rotational
mixing, or diffusion.  Estimates for possible depletion factors are in
the range $\sim$~0.2--0.4~dex \cite{dep}. Recent attempts to deplete
the \li7 abundance through diffusion introduce a source of
turbulence tuned to fit the abundances of heavy elements in NGC6397
\cite{diff}.  It is not clear whether this mechanism will work for the
wide range of stellar parameters seen in the field. As noted above,
the Li data show a negligible intrinsic spread in Li.  Any mechanism
which reduces significantly the abundance of \li7 must do so uniformly
over a wide range of stellar parameters (temperature, surface gravity,
metallicity, rotational velocity etc.).  At the same time, the
mechanism must avoid the nuclear burning of \li7 in order to preserve
some \li6 (which burns at a lower temperature).

In fact, not only does \li6 challenge models of stellar depletion of
\li7, but may require non-standard models to understand its abundance
at low metallicity. Recent data \cite{asp} (though called into
question \cite{cay}) indicates the presence of a plateau in \li6 at
the level of about 1000 times the BBN predicted \li6 abundance
\cite{bbnli6,van}. While this abundance of \li6 can not be explained by
conventional galactic cosmic-ray nucleosynthesis \cite{gcr,fo,van}, it
can be explained by cosmological cosmic-ray nucleosynthesis due to
cosmic rays produced at the epoch of structure formation \cite{ccr}.

It is also possible that the lithium discrepancy(ies) is(are) a sign
of new physics beyond the Standard Model.  One possibility is the
cosmological variation of the fine structure constant. Varying
$\alpha$ would induce a variation in the deuterium binding energy and
could yield a decrease in the predicted abundance of \li7 \cite{vary}.
A potential solution to both lithium problems is particle decay after
BBN which could lower the \li7 abundance and produce some \li6 as well
\cite{jed04}. This has been investigated in the framework of the
constrained minimal supersymmetric Standard Model if the lightest
supersymmetric particle is assumed to be the gravitino \cite{susy} and
indeed, some models have been found which accomplish these goals
\cite{susy2}.

\section{The Lithium Problem Quantified}

\begin{figure}[htb]
\vskip -5cm
\includegraphics[width=18cm]{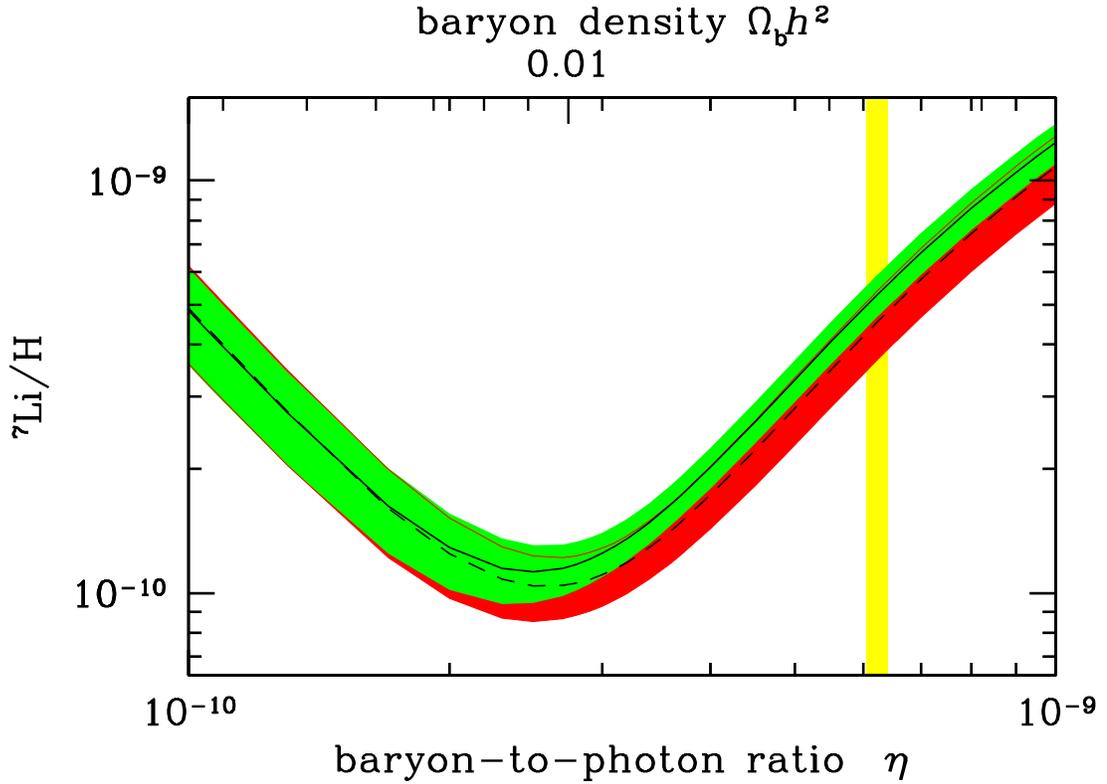}
\vskip -2cm
 \caption{\it A comparison of the \li7/H abundances  as a function of $\eta$
 using the new 
 \he3($\alpha,\gamma$)\be7 rate determined in \protect\cite{cybdav}
 (green band) to the older result from \protect\cite{cyburt} (red band).
 The thickness of the bands represents 1 $\sigma$ uncertainties in the
 calculated abundance.  Central values are given by thick solid and dashed
 curves respectively.  The thin upper curve (within the green shaded region)
 demarcates the border of the old result. The yellow vertical band is the 
 WMAP value of $\eta$ \protect\cite{wmap}. }
  \label{licomp}
\end{figure}

\begin{figure}[htb]
\vskip -5cm
\includegraphics[width=18cm]{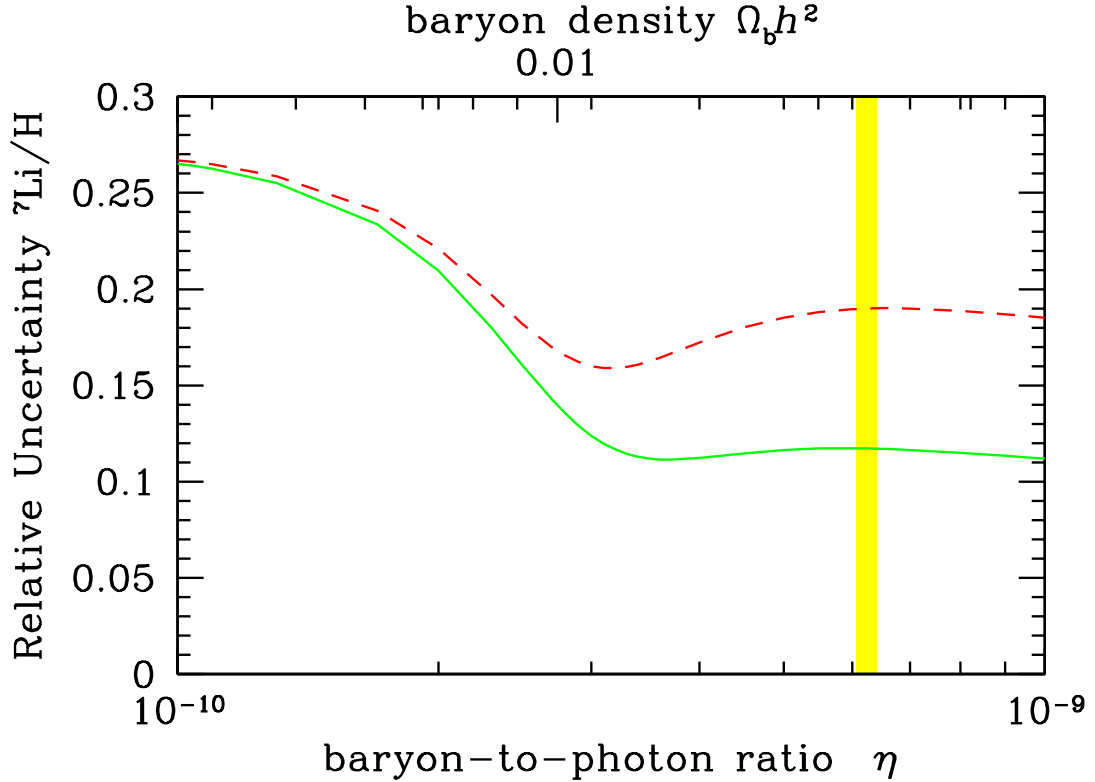}
\vskip -2cm
 \caption{\it The fractional error $\sigma({\rm Li})/{\rm Li}$
in the BBN lithium prediction, shown as a function of $\eta$.
The solid curve is our result; the broken curve
gives the older result from \protect\cite{cyburt} 
as in Fig.~\ref{licomp}.
The yellow vertical band is the 
WMAP value of $\eta$ \protect\cite{wmap}.
}
  \label{fig:relerr}
\end{figure}

As discussed in section IIA, we have incorporated the new cross
section measurements of \he3($\alpha,\gamma$)\be7 as well as an
updated rate for p(n,$\gamma$)d into the BBN calculation of the light
element abundance.  Since the Li problem is our primary focus here, we
first show the effect of the new rate determination for
\he3($\alpha,\gamma$)\be7~\cite{cybdav} in Fig.~\ref{licomp}. There,
the primordial abundance of \li7/H is plotted as a function of $\eta$
or equivalently the baryon density, $\Omega_b h^2$, shown on the upper
horizontal axis. The new result is shown by the green shaded region
which is superimposed on top of the older result taken from
\cite{cyburt} and shown shaded in red. The thickness of the bands
corresponds to one sigma uncertainty in the calculated abundance as
determined by a Monte Carlo simulation of the BBN reaction
network. Also shown as a vertical yellow shaded band, is the one sigma
range for $\eta$ as determined by WMAP \cite{wmap}.  To highlight the
change in the Li error budget, Fig.~\ref{fig:relerr} shows the
fractional uncertainty in Li/H as a function of $\eta$, for the same
two sets of rates as in Fig.~\ref{licomp}.

As expected, at low $\eta$, there is virtually no effect of the new 
rate for \he3($\alpha,\gamma$)\be7, with essentially no change
in the central value
and fractional error of \li7.
However, at higher $\eta$, Fig.~\ref{licomp} shows a 
slight shift to larger values of \li7/H. Most pronounced
at high $\eta$,
however, is the reduction the \li7 fractional error seen in 
Fig.~\ref{fig:relerr}, indicating a greatly
reduced uncertainty in the BBN calculation. At the WMAP value of $\eta_{10}  = 6.23$,
we find 
\beq
{\rm \li7/H} = (5.24^{+0.71}_{-0.62}) \times 10^{-10}
\label{newli}
\eeq
to be compared with the previous value of 
\li7/H = $(4.26^{+0.91}_{-0.86}) \times 10^{-10}$~\cite{cyburt}. 
The shifts in both the central value and the error range
almost entirely reflect the corresponding shifts
in the $\he3(\alpha,\gamma)\be7$ cross section and rate.
 
One may notice the difference between the baryon-density convolved
value of \li7/H=$4.26\times 10^{-10}$ ($\eta_{10}=6.14\pm0.25$) and
the fixed baryon density value of \li7/H=$4.36\times 10^{-10}$
($\eta_{10}\equiv6.14$).  This $\sim2$\% shift is due to $\eta$
dependence of the uncertainties in the \li7/H prediction.  With the
higher precision year 5 WMAP baryon density, this shift is $\la1$\%.
The differences can be fully explained by the simple scalings
from~\cite{cyburt}.

In \cite{coc}, it was found that the adopted prescription for defining
errorbars, underestimated the true uncertainty for the reaction
\he3($\alpha,\gamma$)\be7, due to the visible discrepancy between datasets.  
Therefore, a different method was used to evaluate this reaction's
uncertainty (and only this reaction).  Had this method 
been uniformly applied to all reactions, to account for all
discrepant or non-discrepant data, recognizing that the quoted errors are in
fact minimum (not actual) uncertainties akin to the ``minimal''
errors discussed in~\cite{cfo1},  their prediction would still agree with the result in 
\cite{cyburt}, but with inflated errorbars (\li7/H=$4.15^{+.49}_{-.45}
\times 10^{-10}$).

\begin{figure}[htb]
\vskip -1cm
\includegraphics[width=18cm]{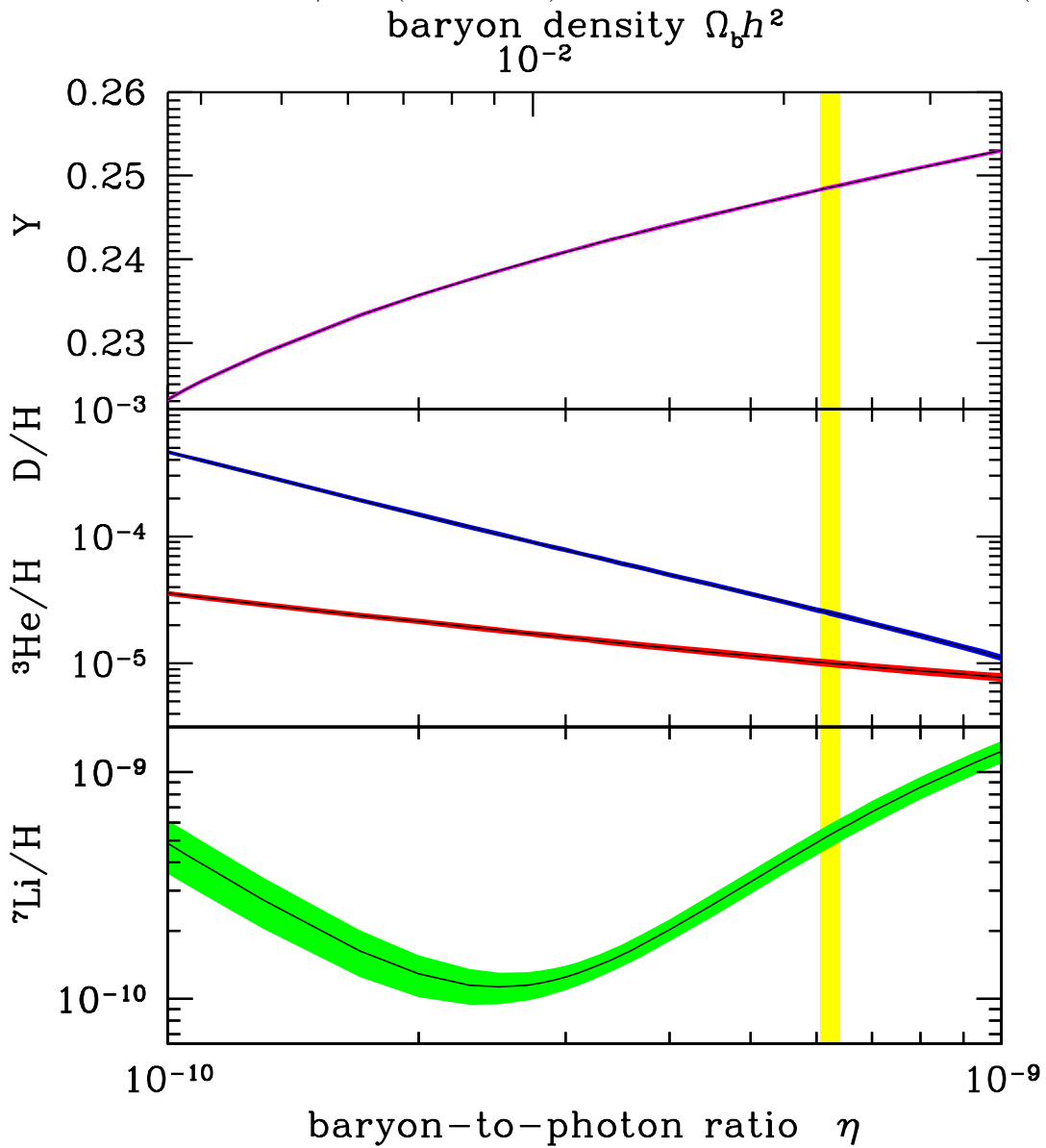}
\vskip -2cm
 \caption{\it The light element abundances of D, \he3, \li7 by number with respect to 
 H, and the mass fraction of \he4 as a function of $\eta$.
 The thickness of the bands represents 1 $\sigma$ uncertainties in the
 calculated abundance.  The 
yellow band gives the WMAP $\eta$ \protect\cite{wmap}. }
  \label{all}
\end{figure}

The full set of light element abundances from BBN is shown in
Fig.~\ref{all}. The \he4 abundance is plotted as a mass fraction as a
function of $\eta$ in the top panel. The thickness of the \he4 band is
primarily due to the small uncertainty in the neutron mean life.  The
D and \he3 abundances by number with respect to H are shown in the
middle panel, and the \li7 abundance (also by number) is shown in the
lower panel.  As one can see, the WMAP determination of the baryon
density is precise enough, that one can now simply read off the
predicted primordial abundances of the light elements.  These are:
\beq
Y_p = 0.2486 \pm 0.0002
\eeq
\beq
{\rm D/H} = (2.49 \pm 0.17) \times 10^{-5}
\eeq
and
\beq
{\rm \he3/H} = (1.00 \pm 0.07) \times 10^{-5}
\eeq

\begin{figure}[htb]
\vskip -1cm
\includegraphics[width=18cm]{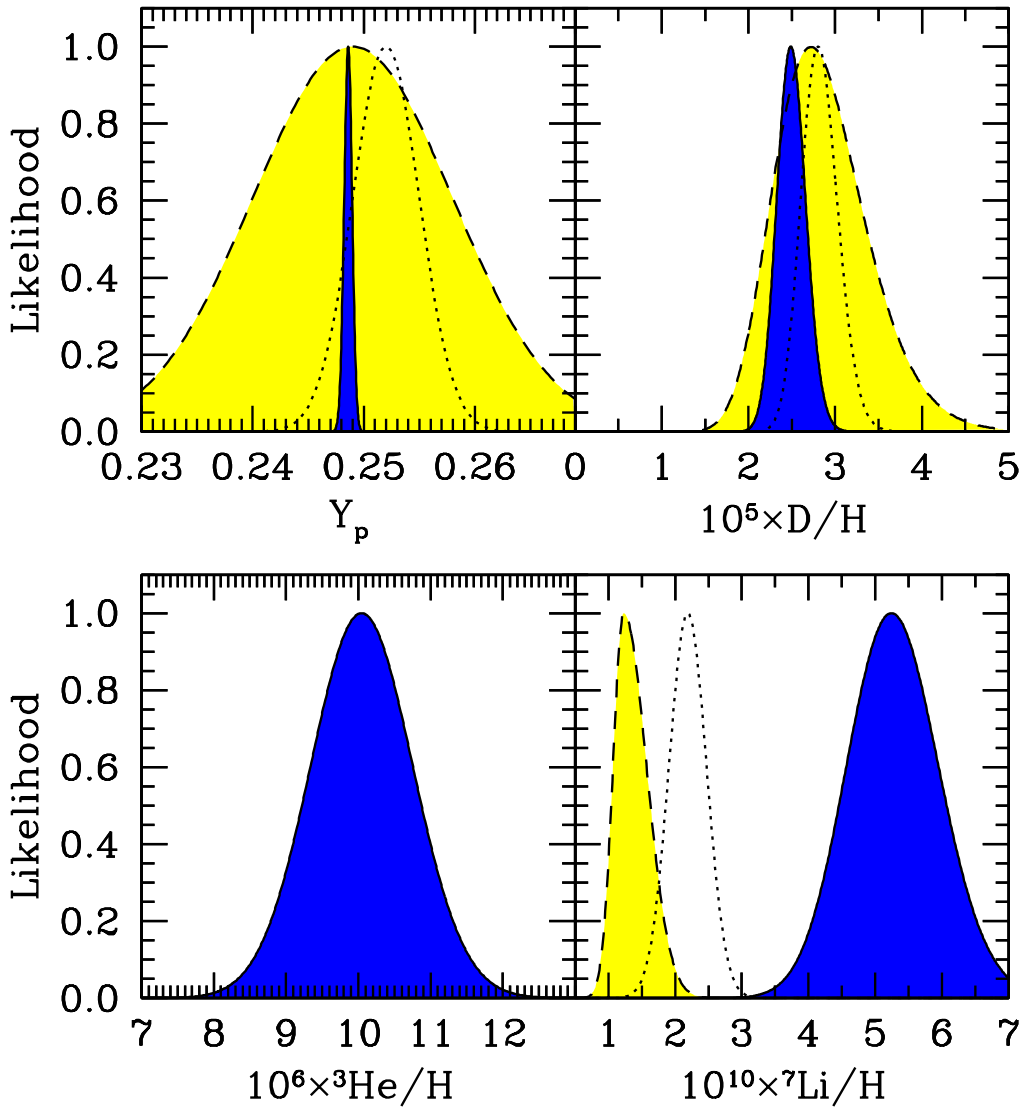}
\vskip -2cm
 \caption{\it The theoretical and observational likelihood functions
 for \he4, D/H, \he3/H, and \li7/H.  BBN results have been convolved with the
WMAP determination of $\eta$ and are  shown as dark (blue) shaded area.
The observational likelihoods are shown as light (yellow) shaded regions as well
as alternative dotted curves. The data and distinctions are detailed in the text.}
  \label{like}
\end{figure}

The BBN predictions can be compared directly with current observational determinations of the
light element abundances. 
The BBN likelihood functions can be defined by a convolution over $\eta$
\beq
L_{\rm BBN}(X) = \int d\eta \ L_{\rm BBN}(\eta | X) \ L_{\rm WMAP}(\eta)
\eeq
using the Monte Carlo results from BBN as a function of $\eta$ to give
$L_{\rm BBN}(\eta | X)$ and the WMAP value of $\eta$ distributed as a
Gaussian, $L_{\rm WMAP}(\eta)$.  These are shown in Fig.~\ref{like} by
the dark (blue) shaded regions.  Though there are useful measurements
of the \he3 abundance \cite{rood}, these are difficult to match to the
primordial abundance \cite{he3}.  We will show the BBN likelihood for
\he3 in Fig.~\ref{like}, but will not discuss \he3 any further.

Fig.~\ref{like} also shows the observational likelihoods for
comparison.  For \he4, the light (yellow) shaded region corresponds to
the result found in \cite{osk} using a select subset of the data in
\cite{it}.  Systematic uncertainties and degeneracies in the set of
physical parameters used to determine the helium abundance led to a
value of $Y$ extrapolated to zero metallicity of $Y_p = 0.249 \pm
0.009$.  The mean value of the reanalyzed data is $0.252 \pm 0.003$
and that is shown by the dotted likelihood function.  Because of the
large observational uncertainty in determining $Y_p$, both agree quite
well with the accurate prediction made by BBN.

The deuterium abundance at low metallicity has been measured in
several quasar absorption systems \cite{pettini}.  The weighted mean value of
the seven systems with reliable abundance determinations is $\log$ D/H
$= -4.55 \pm 0.03$ where the error includes a scale factor of 1.72
\cite{pdg08} and corresponds to D/H = $(2.82 \pm 0.21) \times
10^{-5}$.  Since the D/H shows considerable scatter, it may be
appropriate to derive the uncertainty using sample variance (see
eg. \cite{cfo1}) which gives $\log$ D/H $= -4.55 \pm 0.08$ or D/H =
$(2.82 \pm 0.53) \times 10^{-5}$.  Both are shown plotted as 
log-normal distributions by the dashed curve (light shaded area) and dotted curve 
respectively in the D/H panel of Fig.~\ref{like}. As one can see, there is very
good agreement between the predicted and observationally determined
value of D/H.

Finally, we come to \li7.  The inferred primordial value 
%%%of \li7/H$ = 1.23^{+0.34}_{-0.16} \times 10^{-10}$ from \cite{rbofn} 
from field halo stars (eq.~\ref{eq:field})
is shown by the light shaded area in the Li panel of 
Fig.~\ref{like}.  The slightly higher determination in a globular
cluster 
%%%of \li7/H = $2.19 \pm 0.28$ from \cite{bon1} 
(eq.~\ref{eq:gc})
is shown by the
dotted likelihood function. As one can see, the reduced uncertainty in
the BBN prediction leads to virtually no overlap between the
theoretical and observational likelihoods.

\section{Experimental and Observational Strategies}

We see that the cosmic lithium problem not only remains but has become
even more pressing.  As we have noted already, several proposed
solutions to the Li discrepancy remain viable.  Fortunately, the
different scenarios proposed thus far can be ruled in (or out)
empirically, both in the laboratory and in the observatory.  Here we
outline a strategy for doing so.

The most exciting possibility, in our opinion, is that the lithium
problem points to new physics at work in the early Universe.  This
possibility could receive strong support--in the most optimistic case,
nearly outright confirmation--by the discovery of new physics at the
Large Hadronic Collider (LHC) at CERN, slated to start running in
late 2008.  The presence of particle physics beyond the Standard
Model would immediately cast the Li problem in a new light, since the
correctness of the Standard Model is one of the key assumptions in the
BBN calculation of Li and all of the light element abundances.

Consider, for example, the implications of the LHC finding evidence
for supersymmetry (SUSY). This would place the highest priority on the
study of SUSY effects on nucleosynthesis based on the potentially measurable sparticle
spectrum.  However, the fact that supersymmetry is realized in
nature would not, by itself, guarantee a solution to
the Li problem.  This also requires that SUSY particles alter the light
element abundances either during or after BBN.  The lightest
supersymmetric particle (LSP) would be stable and make up the
weakly-interacting dark matter today, but its feeble interactions
would also make it an inert spectator to nucleosynthesis.  If,
however, the next-to-lightest supersymmetric (NLSP) particle were
relatively long lived ($\tau \sim 1-1000$ sec or more) then its decays
would release a cascade of Standard Model particles which would alter
the light element abundances and potentially solve the Li problem
\cite{jed04,susy,susy2}.  Hence, the most direct laboratory signature
of a SUSY solution to the Li problem would be the discovery of a
long-lived NLSP.

In many of the cases studied \cite{susy,susy2}, the gravitino
is the LSP and the partner of the tau lepton is the NLSP. 
Unfortunately, in the constrained versions of the minimal supersymmetric
Standard Model, if these two sparticles make up the low mass end of the 
spectrum, obtaining lifetimes as short as $\sim 1000$ s, 
would require a very massive spectrum beyond the reach of the LHC.

New astronomical observations of various kinds will play an important
role in further sharpening the Li problem and probing its
solutions.  To date, metal-poor halo stars in our own Galaxy remain
the only observational targets for determining the primordial Li
abundance.  Identifying and exploiting new sites for primitive Li
detection, with different systematic issues, would offer crucial
cross-checks on the halo star results.  For example, metal-poor high
velocity clouds probably represent the Galactic infall of the
primitive intra-group medium of the Local Group. Li abundances in these
clouds, seen in absorption against background quasars, would provide
the first measures of extragalactic Li and with very deep exposures
could even yield isotopic abundances, all without any issues of
stellar evolution or atmospheric modeling \cite{pf}.  An even more
challenging but exciting possibility would be to observe highly
redshifted lines from cosmic Li recombination \cite{zaldarriaga}.
This process occurs after the usual hydrogen recombination but still
at $z \sim 400-500$ and thus the Li features, if measured, would probe
not only the cosmic Li abundance before star formation began, but also
give information about the universe in the ``dark ages.''  Another
observation which bears indirectly on the Li problem will be the new
determination of the extragalactic gamma-ray background by the
now-operational GLAST observatory; these data probe the
non-cosmological but pre-Galactic Li contribution by cosmic rays in
galaxies and on cosmological scales
\cite{ccr}.

If and when new sites for Li determination become available, halo
stars will nevertheless
remain the dominant probe of primordial Li for some time to
come.  Thus there will remain a pressing need to understand
systematics and to obtain abundance measures which are both precise
and accurate.  Further targeted, systematic studies of the kind
outlined in \S \ref{sect:observations} are of the highest importance.
Moreover, robust \li6 data are exceedingly important, as the mere
existence of primordial \li6 at currently observable levels
immediately demands nonstandard nucleosynthesis, and the \li6/H
abundance is a crucial constraint on all models.  Even in the absence
of primordial \li6, the inevitable component from Galactic cosmic-ray
processes, and a possible contribution from cosmological cosmic rays,
both provide unique and empirical means of subtracting the
non-primordial \li7 produced by these mechanisms.

Nuclear experiment and theory remain important inputs to the Li
problem, but as noted above, it would seem that the few relevant
reactions are by now well understood, and a nuclear solution to the Li
problem is unlikely.  That said, it is obvious that the discovery of a
large and unknown systematic error would not only be surprising but
could also dramatically affect the problem.

Finally, we note that these diverse experimental and observational
arenas will contribute in complementary ways.  As LHC probes of SUSY
and other new physics rule particle solutions in or out, astronomical
observations will sharpen the ability of Li and BBN to probe the
solution space.

\section{Conclusions}

The first-year WMAP data determined the cosmic baryon-to-photon ratio
$\eta$ with a high precision that, in concert with BBN theory,
predicts a primordial \li7 abundance significantly above the levels
inferred from the most metal-poor halo stars.  
In this paper we have revisited this cosmic
lithium problem, which has seen important new contributions from
nuclear experiments, halo star observations, and the 5-year WMAP data.
We find that the discrepancy is now a factor 2.4-- 4.3 or $4.2 \sigma$
(from globular cluster stars) to $5.3\sigma$ (from halo field stars),
which is largely due to the tighter errors on the baryon density and
the $\he3(\alpha,\gamma)\be7$ cross section.

A wide variety of possible solutions to the Li problem have been
proposed.  These range from observational systematics in determining
halo star Li, to nuclear effects, to exotic physics beyond the
Standard Model.  Of these, nuclear physics solutions seem essentially
ruled out at this point.  Observational systematics remain a challenging
problem and a possible solution, but the hope that a revised
temperature scale in halo stars would significantly reduce the
Li discrepancy is challenged by recent data which confirm the existing
scale.  The possibility that lithium points to new physics at work in
the early universe thus remains not only viable, but if anything more
likely.

Upcoming experiments and observations will help clarify the lithium
problem and its solution.  In particular, the LHC will probe particle
physics Li solutions generally and supersymmetry in particular.  These
results will be complemented by a host of astronomical observations.
{\em Planck} observations of the CMB, particularly polarization, will
further nail down the cosmic baryon density.  The first extragalactic
Li abundances will offer an independent constraint on primordial Li
with completely different systematics than halo stars. And finally
further halo star measurements, focusing on systematics and on \li6,
will remain crucial in quantifying the Li discrepancy from standard
BBN and probing the mechanism which resolves this disagreement.

R.H.C is supported through the NSF grant PHY 02 16783 (JINA).
The work of KAO
was partially supported by DOE grant DE-FG02-94ER-40823.

\end{document}